\documentclass[preprint, pra,aps,nofootinbib]{revtex4-1}
\usepackage{graphicx,color}
\usepackage{indentfirst}
\usepackage{amsfonts}
\expandafter\let\csname equation*\endcsname\relax
\expandafter\let\csname endequation*\endcsname\relax
\usepackage{amsmath}
\usepackage{amssymb}
\usepackage[T1]{fontenc}
\usepackage{dsfont}
\usepackage{color}
\usepackage{subeqnarray}

\newcommand{\ket}[1]{\left| #1 \right\rangle}
\newcommand{\bra}[1]{\left\langle #1 \right|}

\begin{document}

\title{Small violations of full correlation Bell inequalities for multipartite pure random
states}

\author{R.~C.~Drumond$^{1,2}$ and R.~I.~Oliveira$^{1}$}

\affiliation{$^{1}$ Instituto Nacional de Matem\'{a}tica Pura e Aplicada--IMPA Estrada Dona
Castorina,
110 Jardim Bot\^{a}nico 22460-320, Rio de Janeiro, RJ, Brazil}

\affiliation{$^{2}$ Departamento de Matemática, Universidade Federal de Minas Gerais, caixa postal
702, 30161-970, Belo Horizonte, MG, Brazil}

\begin{abstract}
We estimate the probability of random $N$-qudit pure states violating full-correlation Bell
inequalities with two dichotomic observables per site. These inequalities can show violations that
grow exponentially with $N$, but we prove this is not the typical case. For many-qubit states
the probability to violate any of these inequalities by an amount that grows linearly with $N$ is
vanishingly small.
If each system's Hilbert space dimension is larger than two, on the other hand, the
probability of seeing \emph{any} violation is
already small. For the qubits case we discuss
furthermore the consequences of this result for the probability of seeing arbitrary violations
(\emph{i.e.}, of any order of magnitude) when experimental imperfections are considered.
\end{abstract}

\maketitle

\section{Introduction}

Bell inequality violations substantiate the claim that certain physical phenomena cannot be
described by any local hidden-variables theory. As such, they constitute one of the most striking
"nonlocal" features of quantum mechanics. Besides this foundational interest, Bell inequality
violations also have operational applications in quantum and post quantum information processing.
Example tasks include being able to assure the security of
quantum cryptography~\cite{crip} and reducing the communication complexity of certain
protocols~\cite{caslavcomm}.

It is natural to inquire into the relationship between Bell inequality violations and entanglement,
another nonlocal feature of quantum mechanics. It is known that every pure entangled state violates
some Bell inequality \cite{poproh, dcava}.
A different question is whether Bell inequality violations and entanglement are quantitatively
related. It is known that, in this sense, these two features are genuinely different. For instance,
maximally entangled states do not always achieve maximal violation of Bell's inequalities (see,
\emph{e.g.}, \cite{moreless,anomaly} and references therein).

In this paper we contribute to this quantitative line of research by showing that there is an extreme mismatch between 
entanglement and Bell inequality violations for typical states of $N\gg 1$ qudits. These states are known to
be highly entangled both in the bipartite \cite{entang} and in the multipartite sense
\cite{entanmulti}. By contrast, we will show that the maximal degree of violation of
full correlation Clauser-Horne-Shimony-Holt
 (CHSH)-type inequalities by a typical pure state is very small.  By considering the whole set
of these inequalities, which consist of correlation measurements of dichotomic observables in each
system \cite{zukbru,wol}, we find the following:

(1) If the individual systems are qubits, the maximal violation is typically upper bounded by a
linear function of $N$ (this will mean that we should {\em not} expect to see violations in the
laboratory, as we will argue below).

(2) If the local dimension is strictly larger than $2$, the typical degree of violation is
actually zero.

Thus typical states are highly entangled but do not produce large violations of these Bell
inequalities. This is in sharp contrast with, \emph{e.g.}, generalized Greenberger-Horne-Zeilinger 
(GHZ) states, which are not
very entangled but achieve exponential violation of at least one of the proposed Bell inequalities.
In other words, nearly maximal entanglement across many partitions cannot guarantee strong
violations of Bell inequalities, and vice versa.

Our result connects to other problems that have been recently studied. First, the family of
inequalities we
consider not only is of foundational interest, but also has an information-theoretic property.
Reference~\cite{caslavcomm} shows that a violation of one of these inequalities is a sufficient
condition for
decreasing the complexity of communication protocols. 

Second, our paper adds to the large body of work that addresses how typical a given a physical
property is, when one considers the set of all pure states in a quantum system.
References~\cite{entang,entanmulti} establish that in high-dimensions, most pure states not only
have
entanglement, but almost the maximum of it. The same goes for entanglement dynamics, where most
states follow almost the same trajectory, as far as entanglement is concerned~\cite{entandyn}. The
conditions for a quantum system to reach thermal equilibrium and the definition of a quantum
analogue of the ergodic hypothesis have been approached recently from a typicality perspective, in
the sense of showing that under certain conditions on the system's Hamiltonian and Hilbert space
most
states equilibrate or satisfy
the ergodic hypothesis~\cite{varios, linden}. In the quantum computation field, it was demonstrated
that
most pure quantum states of several qubits are ``useless'' for one-way quantum computation, in the
sense that the computation generated by it is equivalent to one performed by a classical computer assisted by random bits~\cite{comp}.

Finally, we will also argue that the linear violation of Bell inequalities we obtain for typical
many qubits states is noise-sensitive. If we take into account experimental
imperfections, namely, local noise, no matter how small, the majority of states can not violate any
of
the inequalities by a significant amount. This consequence of our result had been noted by Pitowsky
~\cite{pit1}, who conjectured
that the maximal violation of a typical state was of the order $\sqrt{N\ln N}$ (his conclusion still
holds for violations of the order $N$).

\section{The full correlation Bell inequalities}

In general a Bell inequality for $N$
parts can have $n$ distinct measurement settings per site with $m$ distinct outcomes per
measurement. We shall deal with the
case $n=m=2$, \emph{i.e.}, two measurement settings per site, with two outcomes each. Moreover,
we use only full-correlation measurements, where every inequality can be explicitly
written~\cite{zukbru,wol}. If $X=(x_{1},...,x_{N})\in \{0,1\}^{N}$, $A_{0}^{j}=\pm 1$ and
$A_{1}^{j}=\pm1$, $1\leq j\leq N$ represents the deterministic measurement results for the pair of
measurements in site $j$, then the inequalities read~\cite{zukbru}:
\begin{equation}
-1\leq\sum_{X\in\{0,1\}^N}S(X)\prod_{j=1}^{N}\frac{(A_{0}^{j}+(-1)^{x_{j}}A_{1}^{j})}{2}\leq 1,
\end{equation}
where $S:\{0,1\}^{N}\rightarrow\{1,-1\}$. The set of inequalities is given then by all the possible
``sign'' functions $S$, which has a total number of $2^{2^{N}}$. Some of these inequalities are
equivalent, for instance, they are obtained from each other by changing the labels of the outcomes.
Nevertheless, the number of nonequivalent inequalities still grows superexponentially~\cite{wol}.

All these linear inequalities can be replaced though by a single nonlinear
inequality~\cite{zukbru,wol}:
\begin{equation}
\sum_{X\in\{0,1\}^N}|\prod_{j=1}^{N}\frac{(A_{0}^{j}+(-1)^{x_{j}}A_{1}^{j})}{2}|\leq 1,
\end{equation}
and we shall use this inequality for the rest of the paper. For a quantum state $\ket{\psi}$ of $N$
$d$-dimensional systems and a choice of a pair of observables $A^{j}_{x_{j}}$,
for $x_{j}=0$ or $1$, representing measurements with results $\pm1$, for each
system $j=1,...,N$, the nonlinear Bell inequality is evaluated then through the function:
\begin{equation}
 Q_{NL}(\ket{\psi},\mathcal{Q})\equiv\sum_{X\in\{0,1\}^N}|\bra{\psi}\bigotimes_{j=1}^{N}\frac{
(A_ { 0 } ^ { j }
+(-1)^ { x_ { j } } A_ { 1}^{j})}{2}\ket{\psi}|,
\end{equation}
where $\mathcal{Q}$ represents the choices of observables.

\section{Bound for the probability of maximal violation}\label{sec:maxviolation}

It was shown that the so-called generalized GHZ $N$-qubit states $\alpha\ket{0}^{\otimes
N}+\beta\ket{1}^{\otimes N}$, for
$|\alpha|=|\beta|=\frac{1}{\sqrt{2}}$, violate one of these inequalities by an amount of the order
$2^{\frac{N}{2}}$~\cite{mer,wol}. On the other hand, for $N$ odd, and small enough but still greater
than zero $|\alpha|$ (or $|\beta|$), so that the state is still entangled, it can not violate
\emph{any} inequality of this form. 

Considering that we have random pure states drawn according to the normalized uniform
measure on the unit sphere of $(\mathds{C}^{d})^{\otimes N}$, one can still ask: Is it true
that for every $N\geq2$ \emph{almost} all pure states do violate some of these inequalities?
Considering also that for large $N$ most states have almost maximum entanglement (average of
bipartite entanglement over all bipartitions), do they violate some inequality also by a great
amount? 

Our main theorem addresses these questions by
estimating the probability of the event
$\mathcal{A}_{v}=\{\ket{\psi}:\text{sup}_{\mathcal{Q}}Q_{NL}(\psi,\mathcal{Q})> v\}$, which looks to
the maximal violation, optimized over all possible observables, that each state can
achieve.

\textbf{Theorem 1}. \emph{For $N\geq 2$, $d\geq2$ integers, $\ket{\psi}\in(\mathds{C}^d)^{\otimes
N}$ a unit
vector distributed according to the uniform measure in the sphere $S_{2d^{N}-1}$ of}
$(\mathds{C}^d)^{\otimes N}$,
$\mathcal{A}_{v}=\{\ket{\psi}:\text{sup}_{\mathcal{Q}}Q_{NL}(\psi,\mathcal{Q})> v\}$,
\emph{the following inequality holds true:
\begin{equation}
\mathds{P}(\mathcal{A}_{v})\leq
2
\left(\frac{N2^{N+1}d^2}{\delta}+2\right)^{2d^2N}e^{-\frac{(v-\delta-c_{d,N})^{2}(d/2)^{N}}{9\pi^{3
} }},
\end{equation}
for any  $\delta>0$, $v>c_{d,N}+\delta$}, while
$c_{d,N}=(\sqrt{\frac{2}{d}})^{N}+\frac{d-2}{d}$.

An immediate consequence of this bound is that, for large $N$, most pure states do not
get even
close to exhibiting a violation of the order $2^{\frac{N}{2}}$, as generalized GHZ states do. For
qubits we have $c_{2,N}=1$ for every $N$, so as long
as $v\geq cN$, with $c$ a positive suitable constant, we have
$\mathds{P}(\mathcal{A}_{v})\rightarrow 0$ as
$N\rightarrow\infty$. That is, this is a result close to the conjecture presented by
Pitowsky~\cite{pit1}, although the author assumed that a violation already of the order
$\sqrt{N\ln{N}}$ would have vanishing probability.

For $d\geq 3$ we have a more drastic scenario. Here $c_{d,N}\rightarrow \frac{d-2}{d}<1$ with $N$ so
it is
possible to take appropriate $\delta>0$ and $\delta+c_{d,N}<v<1$ such that
$\mathds{P}(\mathcal{A}_{v})$ already
goes to zero (super exponentially). That is, the majority of states do not violate any of these
inequalities if $N$ is large enough. 

\textbf{Idea of the Proof}--The basic idea is to use that $Q_{NL}(\ket{\psi},\mathcal{Q})$ is
``well-behaved'', being
Lipschitz in its variation with $\ket{\psi}$, as well as $\mathcal{Q}$. From this we perform two extra steps:

(1) We may construct an $\epsilon$ net, \emph{i.e}. a discretization for the space of choices for
$\mathcal{Q}$ such that all elements in this continuous space are approximated by an element of the
net up to an error of $\epsilon$. The fact that $Q_{NL}(\ket{\psi},\mathcal{Q})$ is Lipschitz means
that it does not vary for more than $\delta$ in considering the discretized $\mathcal{Q}$, if
$\epsilon$ is chosen accordingly.

(2) Since $Q_{NL}(\ket{\psi},\mathcal{Q})$ is Lipschitz w.r.t $\ket{\psi}$, we upper bound the
probability of
violating the inequality for each $\mathcal{Q}$ in the $\epsilon$-net using L\'{e}vy's
Lemma of measure concentration in high-dimensional spheres~\cite{lev}.

\textbf{Bounding the expected value of $Q_{NL}$}. The first element we need
for the proof is an estimate of the expected value of the function $Q_{NL}$ for a fixed
$\mathcal{Q}$, that is, a fixed choice of measurement observables. To do so we define, for
$j=1,...,N$, $x_{j}=0,1$,
$B_{j,x_{j}}\equiv\frac{1}{2}(A_{0}^{j}+(-1)^{x_{j}}A_{1}^{j})$, and denote by
$\lambda_{i_{j},x_{j}}$,
$i_{j}=1,...,d$ the eigenvalues of $B_{j,x_{j}}$. Since we are concerned with the maximal
violation exhibited by a state, it is enough to consider $A_{0}^{j}$ and $A_{1}^{j}$
unitary Hermitian operators~\cite{werwol2}, \emph{i.e.},
$A_{i}^{j}=(A_{i}^{j})^\dagger=(A_{i}^{j})^{-1}$. Note this includes the operators $I$ and
$-I$.\footnote{It can be indeed advantageous to include these operators, even though they
represent somewhat ``dull'' measurements. Take, for instance, the three-qubit mixed state
$\frac{I}{2}\otimes\ket{\Phi}\bra{\Phi}$, where $\ket{\Phi}$ is any maximally entangled state of
two qubits. If we only perform measurements represented by Pauli operators (\emph{i.e.}, with
eigenvalues $1$ and $-1$), no violation will be seen (all expectation values will be zero). But if
``dull'' measurements are performed on the first qubit, one can see violations due to the
entangled state on the second and third qubits.} It is easy to show then:
\begin{align}
\text{Tr}B_{j,0}^{2}+\text{Tr}B_{j,1}^{2}&=d,\label{eqtr2}\\
|\text{Tr}B_{j,0}|+|\text{Tr}B_{j,1}|&\leq\text{max}\{|\text{Tr}A_{0}^{j}|,|\text{Tr}A_{1}^{j}|\}
\leq d\label{eqtr}.
\end{align}

On the other hand, in order to have a violation of these Bell inequalities we must perform, in at
least one of the systems, a pair of ``nondull'' measurements, \emph{i.e.}, represented by
observables with both eigenvalues $\pm1$. For the pair of observables on this system one can
strengthen inequality \eqref{eqtr}:
\begin{align}
 |\text{Tr}B_{j,0}|+|\text{Tr}B_{j,1}|\leq d-2\label{eqtr'},\text{  for at least one
system $j$.}
\end{align}
To sum up, it is enough to consider measurement settings $\mathcal{Q}$ where each system observable
is Hermitian and unitary and, in at least one of the systems, the pair of observables has both
eigenvalues $\pm1$.

For a fixed $X=(x_{1},...,x_{N})$, we expand an arbitrary
state in the product eigenbasis $\ket{i_{1}}\otimes...\otimes\ket{i_{N}}$ of
$\bigotimes_{j=1}^{N}B_{j,x_{j}}$ [omitting its dependence on $(x_{1},...,x_{N})$ to simplify
notation]. To simplify notation even further, we use capital letters to denote the array
of indexes $(i_{1},...,i_{N})=I$ and write
$\ket{i_{1}}\otimes...\otimes\ket{i_{N}}\equiv\ket{I}$ and
$\lambda_{i_{1}}...\lambda_{i_{N}}\equiv\lambda_{I}$, while $\alpha_{I}$ denote the expansion
coefficient
of a state in this basis. We can write then

\begin{subeqnarray}
\mathds{E}[|\bra{\psi}\bigotimes_{j=1}^{N}B_{j,x_{j}}\ket{\psi}|]&=&\mathds{E}[|\sum_{I}|\alpha_{I}
|^
{
2}\lambda_{I}|]\\
&\leq&
\{\mathds{E}[(\sum_{I}|\alpha_{I}|^{2}\lambda_{I})^{2}]\}^{
1/2}.
\end{subeqnarray}
The inequality is just a particular instance of Jensen's \cite{ineqs}, using that the square root
is a concave function. Noting that
$\mathds{E}[|\alpha_{I}|^4]=2\mathds{E}[|\alpha_{I}|^2|\alpha_{L}^{2}|]=2/d^N(d^N+1)$ for every
$I$ and $L\neq I$,\footnote{In Ref.~\cite{von} the author compute the expected value of any
random variable of the form $(\sum_{I\in\mathcal{I}}|\alpha_{I}|^{2})^{n}$ for any non-negative
integer $n$ and $\mathcal{I}\subset \{0,1\}^{N}$. From these it is straightforward to compute the
expected values we use.} we have:
\begin{subeqnarray}
\{\mathds{E}[(\sum_{I}|\alpha_{I}|^{2}\lambda_{I})^{2}]\}^{
1/2}&=&\{\sum_{I}\mathds{E}[|\alpha_{I}|^{4}]\lambda_{I}^{2}
+\sum_{I\neq L}\mathds{E}[|\alpha_{I}|^{2}|\alpha_{L}|^{
2 }] \lambda_ {I}\lambda_{L}\}^{1/2}\\
&=&\frac{1}{d^{N/2}(d^{N}+1)^{1/2}}\{2\sum_{I}\lambda_{I}^{2}+\sum_{I\neq L}
\lambda_ {I}\lambda_{L}\}^{1/2}\\
&=&\frac{1}{d^{N/2}(d^{N}+1)^{1/2}}\{\sum_{I}\lambda_{I}^{2}+\sum_{I,L}\lambda_{I}\lambda_{L}\}^{1/2
}\\
&=&\frac{1}{d^{N/2}(d^{N}+1)^{1/2}}\{\prod_{j=1}^{N}\text{Tr}B_{j,x_{j}}^{2}+\prod_{j=1}^{N}
(\text {Tr}B_{j,x_{j}})^{2}\}^{1/2}\\
&<&\frac{1}{d^{N}}[\prod_{j=1}^{N}\sqrt{\text{Tr}B_{j,x_{j}}^{2}}+\prod_{j=1}^{N}
|\text {Tr}B_{j,x_{j}}|].
\end{subeqnarray}
In the inequality we use that $(d^{N}+1)^{1/2}>d^{\frac{N}{2}}$ and $(a^{2}+b^{2})^{1/2}\leq
|a|+|b|$ for any $a,b\in \mathds{R}$. Summing over all $X\in \{0,1\}^{N}$, we have
 \begin{align}
\mathds{E}[Q_{NL}(\ket{\psi},\mathcal{Q})]&<
\frac{1}{d^{N}}[\prod_{j=1}^{N}(\sqrt{\text{Tr}B_{j,0}^{2}}+\sqrt{\text{Tr}B_{j,1}^{2}})+\prod_{
j=1}^{N}(|\text {Tr}B_{j,0}|+|\text {Tr}B_{j,1}|)]\\
&\leq\frac{(\sqrt{2}\sqrt{d})^{N}}{d^{N}}+\frac{(d-2)}{d}\equiv c_{d,N}\label{cotave}
\end{align}
while the last inequality is obtained using that $|a|+|b|\leq
\sqrt{2}(a^{2}+b^{2})^{1/2}$ for any $a,b\in \mathds{R}$ and Eqs.~\eqref{eqtr2},~\eqref{eqtr},
and~\eqref{eqtr'}.

\textbf{Bounding the Lipschitz constant of $Q_{NL}$}. Regardless of the local dimension $d$, the
Bell operators
$\mathcal{B}_{S,\mathcal{Q}}=\sum_{X\in\{0,1\}^{N}}S(X)\bigotimes_{j=1}^{N}B_{j,x_{j}}$ satisfy
$||\mathcal{B}_{S,\mathcal{Q}}||_{\infty}\leq 2^{\frac{N-1}{2}}$, where $||\bullet||_{\infty}$
denotes the usual operator norm. Now, given two arbitrary states $\ket{\psi}$
and $\ket{\psi'}$ we have
\begin{align}
|Q_{NL}(\ket{\psi},\mathcal{Q})-& Q_{NL}(\ket{\psi'},\mathcal{Q})|=|\sum_{X\in\{0,1\}^{N}}|\bra{\psi
} \bigotimes_ { j=1 } ^ { N } B_ { j,x_{j}}\ket{\psi}|\nonumber\\
&-\sum_{X\in\{0,1\}^{N}}
|\bra{\psi'}\bigotimes_{j=1}^{N}B_{j,x_{j}}\ket{\psi'}| | \\
&\leq|\sum_{X\in\{0,1\}^{N}}S^{*}(X)[\bra{\psi}\bigotimes_{j=1}^{N}B_{j,x_{j}}\ket{\psi}-\bra{\psi'}
\bigotimes_{j=1}^{N}B_{j,x_{j}}\ket{\psi'}]|\\
&=|\text{Tr}[(\sum_{X}S^{*}(X)\bigotimes_{j=1}^{N}B_{j,x_{j}})(\ket{\psi}\bra{\psi}-\ket{\psi'}\bra{
\psi'
} ) ] |,
\end{align}
where $S^{*}(X)=(\bra{\psi}\bigotimes_{j=1}^{N}B_{j,x_{j}}\ket{\psi}-\bra{\psi'}
\bigotimes_{j=1}^{N}B_{j,x_{j}}\ket{\psi'})/|\bra{\psi}\bigotimes_{j=1}^{
N}B_{j,x_{j}}\ket{\psi}-\bra{ \psi'}\bigotimes_{j=1}^{N}B_{j,x_{j}}\ket{\psi'}|$ if the expression
in the parenthesis is not zero, and (say)
$+1$ otherwise. From von Neumman's trace inequality~\cite{vonineq},
followed by H\"older's \cite{ineqs}, the last expression can be bounded and we get
\begin{align}
 |Q_{NL}(\ket{\psi},\mathcal{Q})-Q_{NL}(\ket{\psi'},\mathcal{Q})|&\leq||\sum_{X}S^{*}(X)\bigotimes_{j=1}^
{ N}B_{j,x_{j}}||_
{\infty}||\ket{\psi}\bra{\psi}-\ket{\psi'}\bra{\psi'}||_{1}\\
&\leq 2^{\frac{N+1}{2}}||\ket{\psi}-\ket{\psi'}||,
\end{align}
where $||\bullet||_{1}$ is the trace operator norm and using that
$||\ket{\psi}\bra{\psi}-\ket{\psi'}\bra{\psi'}||_{1}\leq
2||\ket{\psi}-\ket{\psi'}||$, $||\bullet||$ being just the Hilbert space norm.

\textbf{Variation of $Q_{NL}$ with $\mathcal{Q}$}. Now we estimate how the function $Q_{NL}$
varies when
we change the operators describing the pair of measurements in each site. This will be used
to take a ``representative'' finite subset of the set of measurements (a $\epsilon$ net). First,
observe that
\begin{align}
& ||\bra{\psi}\bigotimes_{j=1}^{N}B_{j,x_{j}}\ket{\psi}|-|\bra{\psi}\bigotimes_{j=1}^{N}
\tilde{B}_{j,x_{j}}\ket{\psi}|
|\leq|\bra{\psi}\bigotimes_{j=1}^{N}B_{j,x_{j}}-\bigotimes_{j=1}^{N}
\tilde{B}_{j,x_{j}}\ket{\psi}|\\
&=|\bra{\psi}[B_{1,x_{1}}\otimes...\otimes B_{N-1,x_{N-1}}\otimes(B_{N,x_{N}}-\tilde{B}_{N,x_{N}})
+\nonumber\\
&B_{1,x_{1}}\otimes...\otimes B_{N-2,x_{N-2}}\otimes(B_{N-1,x_{N-1}}-\tilde{B}_{N-1,x_{N-1}})
\tilde{B}_{N,x_{N}}+\nonumber\\
&...+(B_{1,x_{1}}-\tilde{B}_{1,x_{1}})\otimes
\tilde{B}_{2,x_{2}}\otimes...\otimes\tilde{B}_{N,x_{N}}]\ket{\psi}|\\
&\leq N\text{sup}_{j}||B_{j,x_{j}}-\tilde{B}_{j,x_{j}}||_{\infty},
\end{align}
since $||B_{j,x_{j}}||_{\infty}\leq 1$. Furthermore,
\begin{equation}
 ||B_{j,x_{j}}-\tilde{B}_{j,x_{j}}||_{\infty}=\frac{1}{2}||A_{0}^{j}-\tilde{A}_{0}^{j}+(-1)^{x_{
j}}(A_{1}^{j}-\tilde{A}_{1}^{j})||_ {\infty}\leq
\text{sup}_{i=0,1}||A_{i}^{j}-\tilde{A}_{i}^{j}||_{\infty}.
\end{equation}
Defining
$D(\mathcal{Q},\tilde{\mathcal{Q}})=\text{sup}_{j=1,..,N;i=0,1}||A_{i}^{j}-\tilde{A}_{i}^{j}||_{
\infty}$, we have
\begin{equation}
|Q_{NL}(\ket{\psi},\mathcal{Q})-Q_{NL}(\ket{\psi},\tilde{\mathcal{Q}})|\leq N2^{N}
D(\mathcal{Q},\tilde{\mathcal{Q}}),\label{erede}
\end{equation}
where the factor $2^{N}$ comes from the sum in $X$.

We can cover the set of relevant hermitian operators $A$ in
$\mathds{C}^{d}$ by a parametrization with $d^2$ real numbers
$a_{k}$, corresponding to the real and
imaginary parts of their matrix in a given orthonormal basis. For simplicity, we introduce a norm
$||A||'=\text{sup}_{k=1,...,d^{2}}|a_{k}|$ which satisfies, in particular, 
\begin{equation}
||A||_{\infty}\leq 2d^2\,||A||'.
\end{equation}

Defining the distance
$D'(\mathcal{Q},\tilde{\mathcal{Q}})=\text{sup}_{j=1,..,N;i=0,1}||A_{i}^{j}-\tilde{A}_{i}^{j }||'$,
the set of quantum measurements $\mathcal{Q}$ can be seen as a subset of the hypercube
$[0,1]^{2d^{2}N}$ endowed with the maximum or $\ell_\infty$ norm (the norm of a vector is the
largest absolute value of its coordinates). One can check that for any $\epsilon$ there exists an
$\epsilon$ net $N_\epsilon$ of the hypercube with\footnote{Let $M$ be the largest integer smaller
than $\frac{1}{\epsilon}$ and define
$N_{\epsilon}=\{(\frac{n_{1}}{M+1},...,\frac{n_{2d^{2}N}}{M+1})\in
[0,1]^{2d^2N}:n_{j}=0,1,...,M+1 \text{ for }
j=1,...,2d^{2}N\}$. Clearly, every point of the hypercube is at least
$\frac{1}{M+1}\leq \epsilon$ close to a point of $N_{\epsilon}$ and we have
$|N_{\epsilon}|=(M+2)^{2d^2N}\leq(\frac{1}{\epsilon}+2)^{2d^2N}$.}
\begin{equation}
|N_\epsilon| = \mbox{\# of elements in }N_\epsilon\leq \left(\frac{1}{\epsilon}+2\right)^{2d^2N}.
\end{equation}

Comparing $||A||_\infty$ and $||A||'$ and applying Eq.~\eqref{erede}, we see that we may take
$\epsilon =\delta/d^2N2^{N+1}$ in order to guarantee that two choices of $\mathcal{Q}$ within
distance $\epsilon$ have values of $Q_{NL}$ within distance $\delta$ of each other. This results in
a net with size:
\begin{equation}
|N_\epsilon|\leq \left(\frac{N2^{N+1}d^2}{\delta}+2\right)^{2d^2N}.
\end{equation}

\textbf{Proof of theorem 1}. Finally, with all these elements in hand, we can estimate the
probability of $\mathcal{A}_{v}$:
\begin{align}
\mathds{P}(\text{sup}_{\mathcal{Q}}Q_{NL}&(\ket{\psi},\mathcal{Q})>v)\leq\mathds{P}(\text{
sup}_{\mathcal{Q}\in N_{\epsilon}}Q_{NL}(\ket{\psi},\mathcal{Q})>v-\delta)\label{estimativa0}\\
&\leq\sum_{\mathcal{Q}\in
N_{\epsilon}}\mathds{P}(Q_{NL}(\ket{\psi},\mathcal{Q})>v-\delta)\\
&\leq\sum_{\mathcal{Q}\in
N_{\epsilon}}\mathds{P}(Q_{NL}(\ket{\psi},\mathcal{Q})-\mathds{E}[Q_{NL}(\ket{\psi},
\mathcal { Q}
)]>v-\delta-c_{d,N})\\
&\leq 2|N_{\epsilon}|e^{-\frac{(v-\delta-c_{d,N})^{2}(d/2)^{N}}{9\pi^{3}}}.\label{estimativa}
\end{align}
The third inequality comes from Eq.~\eqref{cotave}, while the last one, assuming $v>\delta+c_{d,N}$,
comes from~\cite{lev, linden}.

\textbf{L\'{e}vy's Lemma.} \emph{For every $\epsilon>0$, $n \geq 1$ integer and $F:S_{n}\rightarrow
\mathds{R}$, a real-valued function with Lipschitz constant $\lambda$ (with respect to the Euclidean
distance),
the following inequality holds true:}
\begin{equation}
\mathds{P}(F-\mathds{E}[F]>\epsilon)\leq 2 e^{-\frac{(n+1)\epsilon^2}{9\pi^3\lambda^2}},
\end{equation}
\emph{where $\mathds{P}$ denotes the uniform probability measure on the sphere $S_{n}$ and $\mathds{E}$ the corresponding expected value.}

\section{Small violations by error-prone qubits}

We have left open the question of whether most (or almost all) qubit states exhibit at least
\emph{some}
violation. Even if that is indeed the case, however, we will show below that these violations would
be extremely sensitive to
experimental errors. This is intuitive, since if each qubit is subjected to some error and each
term of the inequality is a product of measurements on each of them, the overall error will
scale
exponentially, while as we have seen, typically a state exhibits a violation that scales no more
than
linearly.

We prove this formally for a  model of  the inevitable
local noise that each qubit is subjected to. Specifically, we consider the representative case of
local white noise
where each qubit is mapped to
$\rho_{2}\mapsto (1-\lambda)\rho_{2}+\lambda\frac{I}{2}$, where $0\leq\lambda\leq1$. The local
assumption is an
adequate one, since measurements in a Bell-like experiment are especially interesting when the
systems are space like separated, so there must be a time interval
where the systems cannot interact or communicate. The mapping for global pure states is then the following:
\begin{align}
\ket{\psi}\bra{\psi}\mapsto
\rho_{\ket{\psi},\lambda}=\sum_{k=0}^{N}\lambda^{k}(1-\lambda)^{N-k}(\sum_{P\subseteq
\{1,...,N\}:|P^{c}|=k}\text{Tr}_{P^{c}}\ket{\psi}\bra{\psi}\otimes
\frac{I_{P^{c}}}{2^{|P{c}|}}),\label{rolambda}
\end{align}
where $P^{c}$ denotes the complement of $P$. Now the function
$Q_{NL}^{\lambda}(\ket{\psi},\mathcal{Q})$, representing the degree of violation due
to a pure state perturbed by this map is given by
$Q_{NL}^{\lambda}(\ket{\psi},\mathcal{Q})=Q_{NL}(\rho_{\ket{\psi},\lambda},\mathcal{Q})$. To bound
the chances of seeing a violation in this case we just follow the same steps of Theorem 1's proof (see Appendix for details), with the following result:

\textbf{Theorem 2}. \emph{For $N\geq 2$, integer, $\ket{\psi}\in(\mathds{C}^2)^{\otimes
N}$ a unit
vector distributed according to the uniform measure in the sphere $S_{2.2^{N}-1}$ of}
$(\mathds{C}^2)^{\otimes N}$,
$\mathcal{A}_{v}^{\lambda}=\{\ket{\psi}:\text{sup}_{\mathcal{Q}}Q_{NL}^{\lambda}(\psi,\mathcal{Q})> v\}$,
\emph{the following inequality holds true:
\begin{equation}
\mathds{P}(\mathcal{A}_{v}^{\lambda})\leq
2\left(\frac{N2^{N+3}}{\delta}+2\right)^{8N}e^{-\frac{(v-\delta-1)^{2}(2/\chi)^{N}}{9\pi^{3}}},
\end{equation}
for any  $\delta>0$, $v>1+\delta$}, while $\chi=[\lambda+(1-\lambda)\sqrt{2}]^2$.

The point is that, for $\lambda>0$ we have $\chi<2$. Therefore, for any fixed $\delta>0$, the
probability of having a violation larger than $1+\delta$ is vanishingly small for large $N$ due to
the
super exponential factor $\exp{[-\frac{(v-\delta-1)^{2}(2/\chi)^{N}}{9\pi^{3}}]}$.

\section{Conclusion}

We proved that, for large $N$, most $N$-qubit pure states cannot
violate full correlation Bell inequalities by a large amount, when the inequalities are restricted
to two measurements settings per site and two outcomes per measurement. For $N$ $d$-dimensional
systems, with $d\geq 3$, the result is even stronger, where most states do not show any violation.
This constitutes another
instance where there is no quantitative correspondence between Bell inequality
violations and entanglement, since most pure
states are highly entangled.

We have left open the question of whether (necessarily small) violations for qubits are typical. We
argued,
though, that even if they are, small experimental imperfections would make it impossible to actually
probe these violations.

All these results reinforce the insights of Pitowsky in
Refs.~\cite{pit1, pit2}, about why entanglement (and nonlocality, we might add) is never present in
the classical world (this is in addition to the well-known argument based on decoherence). The
author argues that most many-body quantum states and most observables are such that 
the quantum signatures they could reveal are small and henceforth hidden by
experimental imperfections. He bases this line of reasoning on many-qubit states. We add to this  by
showing that if each of the subsystems is ``large,'' which is the case of classical systems, it
gets
even more difficult to see Bell inequality violations. 

Of course, the inequalities we consider form a very restricted set. However, our proof method seems
quite robust. We believe it should be applicable to more general families of
measurements, whenever the total number of degrees of freedom for the measurements is much smaller
than the Hilbert space dimension. The main difficulty for such extensions is a characterization of
maximum violation
inequalities in the spirit of Refs.~\cite{zukbru,wol}.

The authors would like to thank CNPq for financial support and F. G. L. Brand\~{a}o for
useful comments and suggestions.

\appendix
\setcounter{section}{1}
\setcounter{equation}{0}

\section*{Appendix}

Here we show the proof of Theorem 2, which has the same idea and structure as that of Theorem 1.

\textbf{Bounding the expected value of $Q_{NL}^{\lambda}$}. Again, first we need a bound for
$\mathds{E}[Q_{NL}^{\lambda}(\ket{\psi},\mathcal{Q})]$.  To do that we first examine the
contribution of each term in the sum describing $\rho_{\ket{\psi},\lambda}$. For
any $P\subseteq \{1,...,N\}$, we have:
\begin{align}
& |\text{Tr}[\bigotimes_{j=1}^{N}B_{j,x_{j}}(\text{Tr}_{P^{c}}\ket{\psi}\bra{\psi})\otimes
\frac{I_{P^{c}}}{2^{|P^{c}|}}]|\nonumber \\
& =|\text{Tr}_{P}[\bigotimes_{j\in
P}B_{j,x_{j}}(\text{Tr}_{P^{c}}\ket{\psi}\bra{\psi})]\prod_{j \in
P^{c}}\frac{\text{Tr}B_{j,x_{j}}}{2}|\\
&=|\prod_{j \in
P^{c}}\frac{\text{Tr}B_{j,x_{j}}}{2}\text{Tr}[\bigotimes_{j\in
P}B_{j,x_{j}}\otimes I_{P^{c}}\ket{\psi}\bra{\psi}]|\\
& = |\bra{\psi}\bigotimes_{j\in P}B_{j,x_{j}}\bigotimes_{j\in P^{c}}
(\frac{\text{Tr}B_{j,x_{j}}}{2}I_{2})\ket{\psi}|.
\end{align}
The operators $\frac{\text{Tr}B_{j,x_{j}}}{2}I_{2}$ appearing on the tensor product over $P^{c}$ are
either zero or $\pm I_{2}$. Therefore, by the same method we used on the proof of Theorem 1 proof
we can now
bound the expected value:
\begin{align}
&
\sum_{X}\mathds{E}[|\text{Tr}[\bigotimes_{j=1}^{N}B_{j,x_{j}}(\text{Tr}_{P^{c}}\ket{\psi}\bra{\psi}
)\otimes
\frac{I_{P^{c}}}{2^{|P^{c}|}}]|]\nonumber\\
& \leq  \sum_{X}\mathds{E}[|\bra{\psi}\bigotimes_{j\in P}^{N}B_{j,x_{j}}\bigotimes_{j\in P^{c}}
(\frac{\text{Tr}B_{j,x_{j}}}{2}I_{2})\ket{\psi}|] \leq 1.
\end{align}

We have then, using the triangle inequality and the above bound,
\begin{align}
&
\mathds{E}[Q_{NL}^{\lambda}(\ket{\psi},\mathcal{Q})]=\sum_{X}\mathds{E}[|\text{Tr}\bigotimes_{j=1}^{
N}B_{j,x_{j}}\rho_{\ket{\psi},\lambda}|]\\
& \leq \sum_{k=0}^{N}\lambda^{k}(1-\lambda)^{N-k}\sum_{P\subseteq
\{1,...,N\}:|P^{c}|=k}\sum_{X}\mathds{E}[|\text{Tr}[\bigotimes_{j=1}^{N}B_{j,x_{j}}(\text{Tr}_{P^{c}
}\ket{\psi}\bra{\psi})\otimes
\frac{I_{P^{c}}}{2^{|P^{c}|}}]|]\\\
& \leq \sum_{k=0}^{N}\lambda^{k}(1-\lambda)^{N-k}\binom{N}{k}\\
& =1.
\end{align}

\textbf{Bounding the Lipschitz constant of $Q_{NL}^{\lambda}$}. The Lipschitz constant of this
function has a smaller bound than the one we get for $Q_{NL}(\ket{\psi},\mathcal{Q})$ if
$\lambda>0$. This is basically the reason why we get a stronger bound for the probabilities of
seeing small violations in this case.

First we get:
\begin{equation}
|Q_{NL}^{\lambda}(\ket{\psi},\mathcal{Q})-Q_{NL}^{\lambda}(\ket{\psi'},\mathcal{Q})|\leq|\text{Tr}[
(\sum_{X}S^{*}(X)\bigotimes_{j=1}^{N}B_{j,x_{j}})(\rho_{\ket{\psi},\lambda}-\rho_{\ket{\psi'},
\lambda}) ] |,\label{lipzlambda}
\end{equation}
where, similarly, $S^{*}(X)=(\text{Tr}\bigotimes_{j=1}^{N}B_{j,x_{j}}\rho_{\ket{\psi},\lambda}-\text{Tr}
\bigotimes_{j=1}^{N}B_{j,x_{j}}\rho_{\ket{\psi},\lambda})/|\text{Tr}\bigotimes_{j=1}^{N}B_{j,x_{j}}\rho_{\ket{\psi},\lambda}-\text{Tr}
\bigotimes_{j=1}^{N}B_{j,x_{j}}\rho_{\ket{\psi},\lambda}|$ if the expression
in the parenthesis is not zero, and (say)
$+1$ otherwise.

Next we look at the contribution that each term in $\rho_{\ket{\psi},\lambda}$
[Eq.~\eqref{rolambda}] gives to the right-hand side of Eq.~\eqref{lipzlambda}:
\begin{align}
 &|\sum_{X}S^{*}(X)\text{Tr}[\bigotimes_{j=1}^{N}B_{j,x_{j}}\text{Tr}_{P^{c}}(\ket{\psi}\bra{\psi}
-\ket { \psi' } \bra { \psi'})\otimes\frac{I}{2^{|P^{c}|}}]|\nonumber\\
&=|\sum_{x_{j},j\in
P}\text{Tr}_{P}[\bigotimes_{j\in P}B_{j,x_{j}}\text{Tr}_{P^{c}}(\ket{\psi}\bra{\psi}
-\ket { \psi' } \bra { \psi'})]\sum_{x_{j},j\in P^{c}}S(X)\prod_{j\in
P_{c}}\frac{\text{Tr}B_{j,x_{j}}}{2}|\\
& =|\sum_{x_{j},j\in P}c_{x_{j_{1}},..,x_{j_{N-k}}}\text{Tr}_{P}[\bigotimes_{j\in
P}B_{j,x_{j}}\text{Tr}_{P^{c}}(\ket{\psi}\bra{\psi}
-\ket { \psi' } \bra { \psi'})]|,
\end{align}
where $c_{x_{j_{1}},..,x_{j_{N-k}}}\equiv \sum_{x_{j},j\in P^{c}}S(X)\prod_{j\in
P^{c}}\frac{\text{Tr}B_{j,x_{j}}}{2}$, and $j_{i}\in P$ for $i=1,...,N-k$, is such
that $-1\leq c_{x_{j_{1}},..,x_{j_{N-k}}}\leq 1$, since these numbers can be seen as expected values
of full-correlation Bell operators for $k$ qubits on the maximally mixed state and, as such, must
satisfy
the Bell inequality. Therefore, there must exist a sign function $S^{**}:P\rightarrow \{0,1\}^{N-k}$
such that
\begin{align}
& |\sum_{x_{j},j\in P}c_{x_{j_{1}},..,x_{j_{N-k}}}\text{Tr}_{P}[\bigotimes_{j\in
P}B_{j,x_{j}}\text{Tr}_{P^{c}}(\ket{\psi}\bra{\psi}
-\ket { \psi' } \bra { \psi'})]| \nonumber\\
& \leq  |\sum_{x_{j},j\in P}S^{**}(x_{j_{1}},..,x_{j_{N-k}})\text{Tr}_{P}[\bigotimes_{j\in
P}B_{j,x_{j}}\text{Tr}_{P^{c}}(\ket{\psi}\bra{\psi}
-\ket { \psi' } \bra { \psi'})]|,
\end{align}
Hence, we have
\begin{align}
&|\sum_{X}S^{*}(X)\text{Tr}[\bigotimes_{j=1}^{N}B_{j,x_{j}}\text{Tr}_{P^{c}}(\ket{\psi}\bra{\psi}
-\ket { \psi' } \bra { \psi'})\otimes\frac{I}{2^{|P^{c}|}}]|\nonumber\\
&\leq|\text{Tr}_{P}[(\sum_{x_{j},j\in P}S^{**}(x_{j_{1}},..,x_{j_{k}})\bigotimes_{j\in
P}B_{j,x_{j}})(\text{Tr}_{P^{c}}(\ket{\psi}\bra{\psi}
-\ket { \psi' } \bra { \psi'}))]|\\
&\leq ||\sum_{x_{j},j\in P}S^{**}(x_{j_{1}},..,x_{j_{k}})\bigotimes_{j\in
P}B_{j,x_{j}}||_{\infty}||\text{Tr}_{P^{c}}(\ket{\psi}\bra{\psi}
-\ket { \psi' } \bra { \psi'})||_{1}\\
&\leq 2^{\frac{(N-k)-1}{2}}||\text{Tr}_{P^{c}}(\ket{\psi}\bra{\psi}
-\ket { \psi' } \bra { \psi'})||_{1}\\
&\leq 2^{\frac{(N-k)-1}{2}}||\ket{\psi}\bra{\psi}
-\ket { \psi' } \bra { \psi'}||_{1}\\
&\leq 2^{\frac{N-k+1}{2}}||\ket{\psi}-\ket{\psi'}||. \label{boundrolambda}
\end{align}
In the second inequality we use again von Neumman and H\"older inequalities. Realizing that
$\sum_{x_{j},j\in P}S^{**}(x_{j_{1}},..,x_{j_{k}})\bigotimes_{j\in
P}^{N}B_{j,x_{j}}$ is a full-correlation Bell operator for $N-k$ qubits gives us the third. The
fourth is due to the monotone behavior of the trace distance with respect to completely positive
trace-preserving maps (the partial trace being a particular instance of them).

Finally, using the bound \eqref{boundrolambda} and expression \eqref{rolambda} for
$\rho_{\ket{\psi},\lambda}$ we can compute:
\begin{align}
& |Q_{NL}^{\lambda}(\ket{\psi},\mathcal{Q})-Q_{NL}^{\lambda}(\ket{\psi'},\mathcal{Q})|\nonumber\\
& \leq
\sqrt{2}\sum_{k=0}^{N}\lambda^{k}(1-\lambda)^{N-k}\binom{N}{k}\sqrt{2}^{N-k}||\ket{\psi}
-\ket{\psi'} ||\\
&= \sqrt{2}[\lambda+(1-\lambda)\sqrt{2}]^{N}||\ket{\psi}-\ket{\psi'}||.
\end{align}

\textbf{Variation of $Q_{NL}^{\lambda}$ with $\mathcal{Q}$}. The last element we need is an
appropriate $\epsilon-$net to replace the continuous set of measurements by a discrete subset of it.
But this can be obtained in the
exact same way as before and the same bound we get for its number of elements, for $d=2$, apply here
as well.

Finally, following the same line of reasoning of Eqs. \eqref{estimativa0}--\eqref{estimativa} we get
the bound:
\begin{align}
\mathds{P}(\text{sup}_{\mathcal{Q}}Q_{NL}^{\lambda}&(\ket{\psi},\mathcal{Q})>v)\leq
2(\frac{N2^{N+3}}{\delta}+2)^{8N}e^{-\frac{(v-\delta-1)^{2}(\frac{2}{\chi})^{N}}{9\pi^{3}}},
\end{align}
where $\chi=(\sqrt{2}-\lambda(\sqrt{2}-1))^2$.

\end{document}